\documentclass[a4paper]{article}
\usepackage{subcaption}

\usepackage{times}
\usepackage{lscape}
\usepackage{bm}
\usepackage{natbib}
\usepackage{amsmath}
\usepackage[figuresright]{rotating}
\usepackage{graphicx}
\usepackage{amssymb}
\usepackage{amsmath}
\newcommand{\NN}{{\cal N}}

\newcommand{\intr}{{\rm int}}

\newcommand{\Ee}{{{\rm E}}}

\newcommand{\bX}{\mbox{\boldmath $X$}}

\newcommand{\dd}{{\rm d}}

\newcommand{\dinf}[1]{\longrightarrow_{#1}}

\newcommand{\Ob}{{{\cal O}}}

\newcommand{\PP}{{{\cal P}}}

\newcommand{\FF}{{{\cal F}}}

\usepackage{colortbl}

\title{Effects of interventions and optimal strategies in the stochastic system approach to causality}

\author{Daniel Commenges and M\'elanie Prague\\
INSERM, U 1219, Bordeaux,  F33076, France}

\begin{document}





\label{firstpage}

\maketitle
\begin{abstract}
We consider the problem of defining the effect of an intervention on a time-varying risk factor or treatment for a disease or a physiological marker; we develop here the latter case. So, the system considered is $(Y,A,C)$, where $Y=(Y_t)$, is the marker process of interest, $A=A_t$ the treatment (assumed to take values $0$ or $1$) and $C$ a potential confounding factor. The marker process $Y$ has a Doob-Meyer decomposition $\dd Y_t=\lambda_t \dd t+\dd M_t$, where the intensity of the process $Y$, $\lambda_t$ is a function of the past history of the three processes and can be written as $\phi(\bar Y_{t-},\bar A_{t-},C))$, where $\bar X_t$ means the information of $X$ up to time $t$; the function $\phi(\cdot, \cdot,\cdot)$ is the ``physical law'' and cannot be changed. $Y$ lives in continuous time but can be observed only at discrete times by: $Z_{j}=Y_{t_j}+\varepsilon_j$. A realistic case is that the treatment can be changed only at discrete times, according to a probability law: $\PP(A_{t_j}=1| \bar Z_{j}, \bar A_{t_{j-1}},C)$. In an observation study the treatment attribution law is unknown; however, the physical law can be estimated without knowing the treatment attribution law, provided a well specified model is available. An intervention is specified by the treatment attribution law, which is thus known. Simple interventions will simply randomize the attribution of the treatment; interventions that take into account the past history will be called ``strategies''. The effect of interventions can be defined by a risk function  $R^{\intr}=\Ee_{\intr}[L(\bar Y_{t_J}, \bar A_{t_{J}},C)]$, where
$L(\bar Y_{t_J}, \bar A_{t_{J}},C)$ is a loss function, and contrasts between risk functions for different strategies can be formed. Simple contrasts between two strategies, like $\Ee_{\intr 1}(Y_{t_J})-\Ee_{\intr 0}(Y_{t_J})$, are very particular cases of this approach. Once we can compute effects for any strategy, we can search for optimal or sub-optimal strategies; in particular we can find optimal parametric strategies. We present several ways for designing strategies. As an illustration,  we consider the choice of a strategy for containing the HIV load below a certain level while limiting the treatment burden. A simulation study demonstrates the possibility of finding optimal parametric strategies.
\end{abstract}


\section{Introduction}\label{sec:Introduction-DynCaus}
It is of great importance in epidemiology and public health to identify and quantify causal effects of a factor on the risks of diseases and death. The ultimate aim of such research is to decrease such risks by modifying the factors that can be modified. Factors that can be modified, although not always easily and completely, may be treatments, exposures, life-style. Modification will be realized through an intervention. It will be possible to compute the effects of interventions if, and only if,  the causal effects have been correctly estimated. Having understood the difficulty of establishing it, epidemiologist have long be reluctant to speak of causality, although \citet{Hill1965} stated a list of common sense criteria which are still useful. More recently, statisticians developed formalisms for causality; two broad approaches can be distinguished: the ``potential outcome'' and the ``dynamic'' approaches.

 The use of potential outcomes has first been proposed by Jerzy Neyman in 1923 \citep{Splawa-Neyman1990} but was formulated in modern notation much later by \citet{rubin1974estimating}. Since then, causal inference based on potential outcomes have been developed in a series of papers and has become the dominant school of causal inference in biostatistics and has also been influential in other fields such as econometrics \citep{Heckman2005a}. This theory was reviewed by \citet{holland1986statistics} and \citet{rubin2005causal} among others. However, the potential outcome approach has been criticized \citep{Dawid2000,commenges2019causality}.

 Another approach relies on developing dynamical models. This is in line with the definition of \citet{Granger1969} working with time series and has been given a more powerful formalism by the works of \citet{Schweder1970} and \citet{Aalen1989linear}, further developed by \citet{arjas2004causal} and \citet{Didelez2008}. Of course, the use of a dynamical model is not sufficient for causal interpretation but it is possible to formalize the assumption needed, particularly through the concept of ``system'' elaborated in \citet{Commenges2009}, in Chapter 9 of \citet{commenges2015dynamical} and in \citet{commenges2018dealing}; and this was called ``the stochastic system approach to causality''.

One issue that has retained much attention recently is that the factor of interest, that will be called ``treatment'' in the following, may be dynamic and maybe influenced by factors linked to the outcome. Robins and coworkers used the potential outcomes approach to these complex problems \citep{Robins2000,robins2009estimation}. It was shown that these problems can also be tackled through the dynamic approach \citep{arjas2012causal,prague2016dynamic}. Finally, when the causal effect has been estimated, one can compute the effects of interventions, and one can further try to find the best interventions. These interventions can be personalized and adaptive, leading to the search of optimal treatment regime. This topic has attracted much attention recently \citep{murphy2003optimal,chakraborty2014dynamic,saarela2015predictive,shen2017estimation,hager2018optimal}.

 One aim of this paper is to show that within the stochastic system approach to causality, the effects of interventions can be estimated from observational studies without resorting to counterfactuals or potential outcomes; this can be done either in a direct way or through the use of marginal structural models. Thus the effects of any complex interventions can be estimated in a way similar to \citet{Pearl2000} but with the important difference that our approach is based on stochastic processes rather than random variables. The other aim is to show that optimal and sub-optimal strategies can be designed and estimated. Our approach incorporates realistic modeling of important features, acknowledging that biological processes live in continuous time while observations are made at discrete times and with errors; see \citet{aalen2016can} for the importance of continuous time.

 In Section \ref{sec:context} we recall the main features of the stochastic system approach to causality and we introduce a lead example  that will be used to illustrate the theory. In Section \ref{condlawknown}, we define the effects of interventions, we see how to compute them and in Section \ref{sec:strategies} we examine some ways of designing strategies. Since finding optimal strategies is in general impractical, we propose parametric families of strategies.
Then in Section \ref{condlawunknown}, we examine the inference issue, showing that we can estimate the physical law directly and then compute the effect of any intervention. Once we can compute effects for any strategy, we can search for optimal strategies. In Section \ref{sec:illustration}, we present a simulation study illustrating the choice of a strategy for containing the HIV load  below a certain level while limiting the treatment burden. We then conclude.

\section{The context}\label{sec:context}
\subsection{Recall of the Stochastic System Approach to Causality}\label{sec:recall}
In the framework of the stochastic system approach to causality, we consider a (voluntarily simple) system $\bX=(Y,A,C)$, where $Y,A,C$ are counting processes or diffusion processes, $Y$ is the outcome of interest, $A$ a treatment, $C$ an observed {\em potential} confounding factor; $C$ is really a confounding factor in an observational study; note that that the property of ``confounding'' depends on both the system and the law \citep{commenges2018dealing}. Influences between processes are defined via ``local independence'' \citep{Aalen1987}, direct influence being the contrary of local independence.
Given a system represented by a multivariate stochastic process $\bX$ which may include both counting and diffusion processes, a criterion of local independence  is defined in terms of measurability of processes involved in the Doob-Meyer representation. \citet{Commenges2009} denoted the local independence by WCLI (weak local conditional independence) because they also defined a criterion of strong local independence (SCLI); when WCLI does not hold, there is direct influence, when SCLI does not hold while WCLI holds, there is indirect influence. In short, if a component of the stochastic process $X_k$ does not appear in the compensator of the Doob-Meyer decomposition of $X_j$ we say that $X_j$ is WCLI of $X_k$.
A system $\bX$ is called ``perfect'' for $Y$ if we cannot find a process $U$ which influences $Y$ in the augmented system  $\bX'=(Y,A,C,U)$. A system is ``NUC'' (no unmeasured confounder) for the effect of $A$ on $Y$ (in short for $[A \rightarrow Y]$) if there does not exist a process $U \not\in \bX$ such that $U \dinf{\bX'} Y$ and $U \dinf{\bX'} V$; see \citet{commenges2018dealing}.

In this paper, we will assume that both systems $\bX$ and $\bX'=(Y,A,C,U)$ are NUC for $[A \rightarrow Y]$. These assumptions allow interpreting influences as causal influences; in $\bX'$, $U$ will be treated as a random effect (see Section \ref{sec:observation}). If $X$ is perfect, conditional and marginal effects of $A$ on $Y$ can be estimated \citep{Commenges2009}; if $X$ is NUC for $[A \rightarrow Y]$, conditional effect of $A$ on $Y$ with respect to $C$ but marginal with respect to other factors influencing $Y$, can be estimated \citep{commenges2015dynamical,commenges2018dealing}.  To the system $\bX$ is associated the filtration $\FF=(\FF_t)$, where $\FF_t$ is the sigma-field generated by $(\bar Y_t, \bar A_t, \bar C_t)$, noted $\FF_t=\sigma(\bar Y_t, \bar A_t, \bar C_t)$. We use the notation $\bar Y_t=(Y_u, u\le t)$, and similarly for the other processes. Similarly, to the system $\bX'$ is associated the filtration $\FF'$.

\subsection{Physical law and treatment regime}\label{sec:Physlaw}

\subsubsection{Physical law}
As in \citet{commenges2015stochastic}, it is important to define the ``physical law'', allowing for heterogeneity between subjects; here it is necessary to restore the subscript ``i''. We assume that the Doob-Meyer decomposition of $Y_i$ in $\FF^{'i}$ can be written:
\begin{equation}
    Y_{it}=\phi_t(\bar Y_{it-},\bar A_{it-},C_i,U_i)+M_{it},
\end{equation}
where $\phi_t(\cdot,\cdot,\cdot,\cdot)$ is a function which does not depend on $i$ and the martingales are orthogonal. The conjunction of $\phi$ and the law of the martingales define the physical law of $Y$.

\subsubsection{Observation}\label{sec:observation}
We must acknowledge that generally we do not completely observe $Y_{it}$ for all $t$. For instance if $Y_i$ represents an event, its observation may be right-censored and if $Y_i$ is a quantitative marker, the observation is made at discrete times and with a measurement error. We denote by $\bar Z_{it}$ the observation of $Y_i$ up to time time $t$. Acknowledging this fact is important for both inference and for defining the treatment regime. We assume that both $A_i$ and $C_i$ are exactly observed. Thus the observation up to time $t$ is the sigma-field $\Ob_{it}=(\bar Z_{it},\bar A_{it},C_i)$. The family of these sigma-fields is the filtration $\Ob_i=(\Ob_{it})$. $U_i$ is not observed and can be considered as a random effect.    

A more complex observation scheme arise if the visit-times are also random; their law may be described by a counting process $N$ (jumping at visit times), the compensator of which could depend on the past of $Y, Z, A, C$. If we restrict to cases where it depends only on past values of  $Z, A, C$, this mechanism is ignorable for inference \citep{commenges2018dealing}. Adapting visit-times may lead to efficient strategies \citep{Villain2018}, but we do not develop this possibility in this paper.

\subsubsection{Treatment regime}
Often, there are visits at which $Y_i$ is observed and the treatment can be changed just after the observation and is unchanged between visits. So, $A_i$ is a process which may change at times $t_{ij}$ and which may be influenced by $Y_i$ through the observations $\bar Z_{ij}$. Rather than writing a Doob-Meyer decomposition in $\Ob_i$ for the process $A_i$, we write the distribution of $A_{it_{ij}}$.Taking the case where the treatment takes a binary value we have:
\begin{equation}
    \PP (A_{it_{ij}}=1|\bar Z_{ij},\bar A_{it_{ij}-1},C_i)=\psi_{it_{ij}}(\bar Z_{iij},\bar A_{it_{ij}-1},C_i).
\end{equation}
The treatment regime is defined by the function $\psi_{it_{ij}}(\cdot, \cdot,\cdot)$. Note that it may depend on $i$. Treatment regimes may vary between subjects because different doctors may have applied different protocols. In observational studies, the treatment is given preferentially  when the observed marker indicates a degradation of the health of the patient: if we wish to estimate the effect of $A$ by naive methods we are confronted to ``dynamic confounding'', or ``indication bias''. Moreover, we do not generally know the treatment regime. On the other hand, we may wish to find an optimal treatment regime (or strategy) for future patients; in that case, there is no reason that $\psi$ depend on $i$. That is, the strategy specified by $\psi_{t_{ij}}(\cdot, \cdot,\cdot)$ does not depend on $i$ although the decision depends on all the information $(\bar Z_{ij},\bar A_{it_{ij}-1},C_i)$ we have on subject $i$ at time $t_{ij}$.

\subsection{Lead toy example}\label{sec:LeadEx}
We shall use a lead toy example to make all the developments more concrete. Let $Y$ be a quantitative marker which may represent the log of the viral load of HIV infected subjects, $C$ may represent infection by intravenous  drug-use and $A$ an antiretroviral treatment. 
Assume that, under the true probability, $Y_i$ has a  Doob-Meyer decomposition in $\FF^{'i}$ which in differential form is:
\begin{equation} \dd Y_{it}=(\mu_{1i}^* + \gamma^*_C C_i+ \gamma^*_A A_{it})\dd t + \tau^*\dd B_{it} ~~;~~ t\in (0,t_f), \label{eq:diffDoobidv}, \end{equation}
with a normality assumption for $\mu_{1i}^*=\mu_{1}^*+U_i$,
and where $\mu_1^*, \gamma^*_C, \gamma^*_A, \tau^*$ are true values; $\gamma^*_A$ should be negative for the treatment to be efficient, and the $B_i$ are standard Brownian motions.
Heterogeneity is explained by the variable $C_i$ but there can also be unexplained heterogeneity that can be represented by random effects on the baseline slope ($\mu_{1i}^*$) and on the initial condition $Y_{i0}=\mu_{0i}$.

Assuming for simplicity that the number of visit-times is the same for all subjects, the observation are:
\begin{equation} Z_{ij}=Y_{it_{ij}}+ \varepsilon_{ij}; j=1,J, \label{eq:obs}\end{equation} with the $\varepsilon_{ij} $ independent normal variables with zero mean and variance $\sigma^{*2}_{\varepsilon}$.

\begin{figure}[h!]

\centering
\includegraphics[scale=0.45]{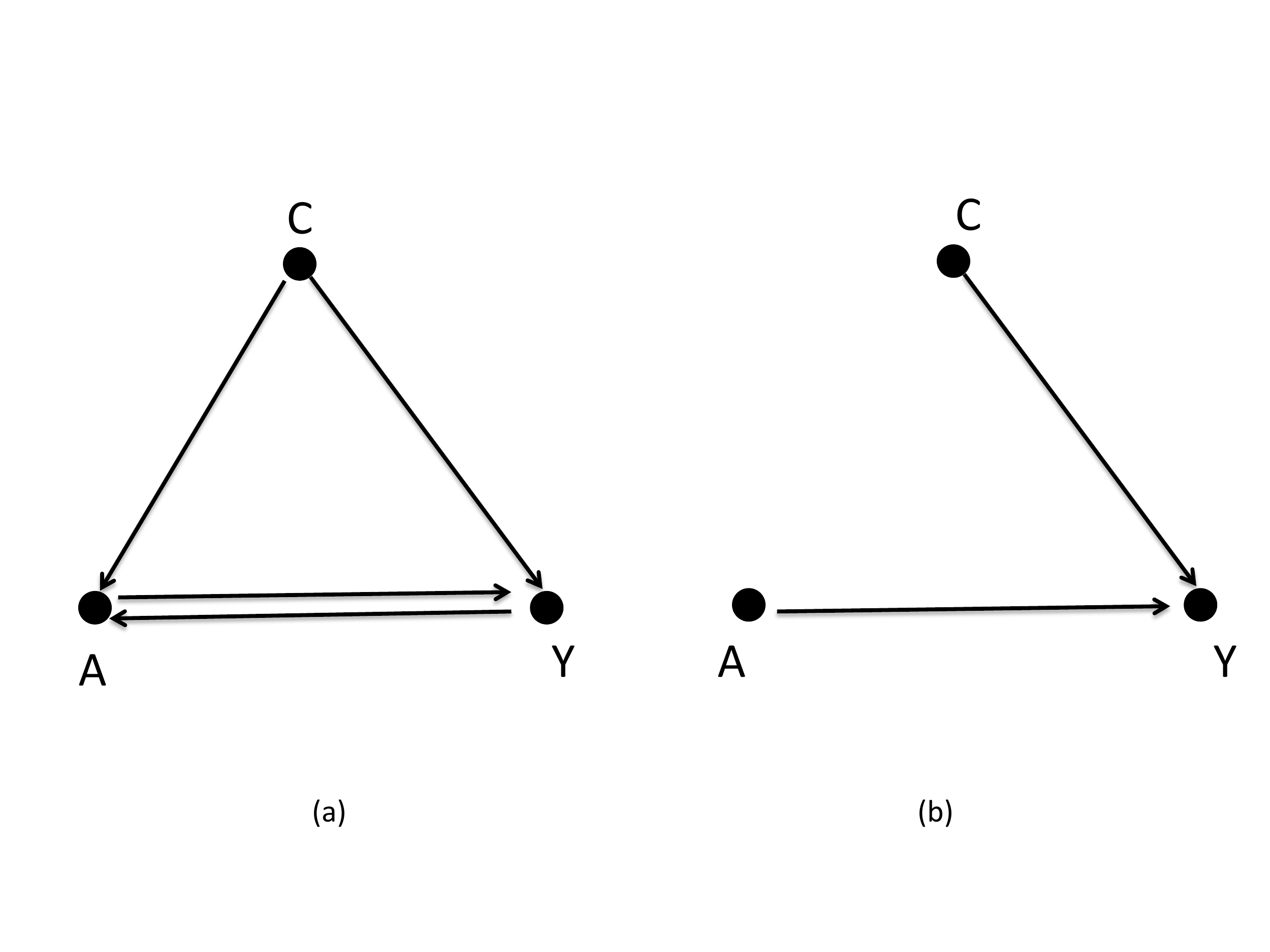}
\caption{Influence graphs for $A$, $Y$ and $C$: (a) observational case or strategy (see text); (b) simple randomized experiment.\label{graph-causal-obs}}
\end{figure}

In order to generate data for a simulation study (see Section \ref{sec:illustration}), we will have to specify  the treatment regime.
We will consider the case where  the probability  of being treated at $t_j$ depends on the observation $Z_{ij}$ and on the factor $C_i$:
\begin{equation} {\rm logit} [\PP(A_{it_{ij}}=1|\bar Z_{ij},\bar A_{it_{ij}-1}, C_i)] =\alpha^*_0+\alpha^*_Z Z_{ij}+\alpha^*_C C_i+\alpha^*_A A_{it_{ij}-1}.\label{eq:strat}\end{equation}
If $\alpha^*_A$ takes a large value, the treatment once given cannot be removed, the case considered by \citet{Hernan2002}. There are, however, active research for the possibility of intermittent treatment \citep{lau2019clinical}.
The influence graph of this system in the observed probability measure \citep{Commenges2009} is depicted in Figure \ref{graph-causal-obs} (a).


\subsubsection{Interventions, experiments, strategies}

How can we use the knowledge that we can obtain about the physical law of $Y$? The scientific context is that we assume that the compensator of $Y$ represents a ``physical law'' which is stable (cannot be changed), while the treatment regime, that is the law of $A$, can be manipulated. Thus, we assume that we can devise strategies which define a probability measure $\PP^{\intr}$ for which the compensator of $Y$ is unchanged but the law of $A$ can be controlled, subject to the constraint that $A$ must remain adapted to the filtration $(\Ob)$. That is, the dynamics of $A$ may only depend on past observed values of $Y,A,C$ (that is of past values of $Z,A,C$). Thus, assuming the observations times $t_{ij}$ fixed, a strategy is specified by  functions $\psi_{t_{ij}}(\cdot,\cdot,\cdot)$ such that:
\begin{equation}
    \PP (A_{it_{ij}}=1|\bar Z_{ij},\bar A_{it_{ij}-1},C_i)=\psi_{t_{ij}}(\bar Z_{ij},\bar A_{it_{ij}-1},C_i).
\end{equation}
In most cases there is not need to make $\psi$ depend on $t_{ij}$.
The terms ``intervention'', ``policy'', ``strategy'' are often used for naming a control which may depend on past information.  In this paper, we use the general term of ``intervention'' to mean any type of control of the law of $A$, and we reserve the term ``strategy'' for adaptive interventions, that is, when the law of $A$ is made dependent of the past observations of $Y$ and $C$. Thus the influence graph under a strategy is generally the same as that in an observational study; the difference is that, for a strategy, the dynamics of $A$ is known.

A simple randomized experiment is a particular case of intervention, where the dynamics of $A$ is made independent of both $Y$ and $C$ (the simplest one is when $A$ takes a randomly chosen constant value); in that case the influence graph is as in Figure \ref{graph-causal-obs}b, analogous to the application of Pearl's ``do-operator'' \citep{Pearl2000}. We assume that we cannot manipulate $C$ which has its own independent dynamics; however we could still consider situations where $C$ would have a different dynamics (or a different distribution in the case where $C$ is a simple random variable), for instance if we apply the model to a different population.

\section{Risk functions}\label{condlawknown}
In this section (Section \ref{condlawknown}), we drop the index $i$ for simplicity of notation, and we work with $\bX$ rather than $\bX'$.
\subsection{Assessing whether there is a causal effect}

The first conclusion that we can derive from the knowledge of the $\FF$-compensator of $Y$ is that if $\lambda^{\FF}_Y \ne \lambda^{\FF'}_Y$, then $A$ influences $Y$. In our example this can of course be seen easily by looking at Equation (\ref{eq:diffDoobidv}). If $\gamma_A^*\ne 0$, $A$ influences $Y$ ($A\dinf {\bX} Y$), and if the system is NUC for $[A \rightarrow Y]$, this influence is causal. This is a conclusion of explanatory nature.

\subsection{Definition of effects as risk functions}
\subsubsection{Generic effects}\label{sec:geneffect}

If there is a causal influence of $A$ on $Y$ we wish to quantify it. The effect of a treatment regime can be defined as a contrast between risk functions or utility functions of this regime and a reference regime. A risk function is the expectation of a loss function;  minus a loss function is a utility function and its expectation is the expected utility; the two formulations are equivalent; here we adopt the formulation in terms of risk function.  For instance, if our loss function is  $Y_{t_J}$, taking the difference of risk  functions as the contrast, we can define an effect as
$$\Ee_{\intr 1}(Y_{t_J})-\Ee_{\intr 0}(Y_{t_J}),$$ where in intervention ``int1'' the treatment is given during the whole period while in intervention ``int0'', the treatment is never given.
This kind of effect is often estimated in simple randomized trials where there is a ``treatment arm'' and a ``placebo arm''. However, a much more general definition is possible, and this will be useful for designing strategies (see Section \ref{sec:strategies}).

\subsubsection{Effects of strategies: risk functions}
 A general loss function is $L(\bar Y_{t_J}, \bar A_{t_{J}},C)$: in general , the loss function reflects a compromise between improving the trajectory of $Y$ and reducing the burden of the therapeutic protocol (treatment dose and number of visits). Risk functions of an intervention are expectations of the loss function in the probability defined by the intervention ``int''.
\begin{itemize}
\item The marginal risk is $R_m^{\intr}=\Ee_{\intr}[L(\bar Y_{t_J}, \bar A_{t_{J}},C)]$;
\item The conditional risks: $R_c^{\intr}(C)=\Ee_{\intr}[L(\bar Y_{t_J}, \bar A_{t_{J}},C)|C]$
\end{itemize}
It is clear that $R_m^{\intr}=\Ee_{\intr}[R_c^{\intr}(C)]$. Note that even if the loss function does not depend on $C$, marginal and conditional risks are in general different. Also, optimizing $R_c^{\intr}(C)$ for all $C$ leads to optimizing $R_m^{\intr}$.

Examples of simple loss functions are $L(\bar Y_{t_{J}}, \bar A_{t_{J}},C)=Y_{t_{J}}$ and $L(\bar Y_{t_{J}}, \bar A_{t_{J}},C)=1_{Y_{t_{J}}>\eta}$, in which case the risk is $R_m^{\intr}=\Ee_{\intr}[Y_{t_{J}}]$ and $R_m^{\intr}=\PP^{\intr}[Y_{t_{J}}>\eta]$, respectively. More complicated loss functions could include values of the whole trajectory of $Y$ and a cost of treatment (both financial and related to toxicity and quality of life). 

A rather general additive risk function for the marker and the treatment is:
\begin{equation}
   R_m^{\intr}=\frac{1}{t_J} \left[\int_0^{t_J} \Ee^{\intr}[h(Y_u)]\dd G(u) + \int_0^{t_J}\Ee^{\intr}[g(A_u)] \dd u\right]
\end{equation}

Particular cases are obtained by taking $G(x)=x$, $g(x)=\omega x$ and $h(x)=x$ or $h(x)=1_{x>\eta}$:

\begin{equation}\label{additiverisk-moyenne} R_m^{\intr}=\frac{1}{t_J} \left[\int_0^{t_J}\Ee^{\intr}[Y_u]\dd u+\omega \int_0^{t_J} \Ee_{\intr}(A_u) \dd u.\right]\end{equation}
or
\begin{equation}\label{additiverisk-seuil} R_m^{\intr}=\frac{1}{t_J} \left[\int_0^{t_J}\PP^{\intr}[Y_u>\eta]\dd u+\omega \int_0^{t_J} \Ee_{\intr}(A_u) \dd u.\right]\end{equation}

Two interventions can be compared by contrasting their risk functions; it is of course interesting to find the best strategies (see Section \ref{sec:strategies}).

\subsection{Examples}\label{sec:examples}
 In our toy example, because of the linearity of the compensator in (\ref{eq:diffDoobidv}), the marginal and conditional (additive) generic effects are the same. For instance the additive contrast (\ref{sec:geneffect}) $\Ee_{\intr 1}(Y_{t_J})-\Ee_{\intr 0}(Y_{t_J})$ is $\gamma^*_A t_J$. Beware that this does not mean that we can estimate these effects with a marginal regression model.

On the other hand, conditional and marginal effects are in general different for strategies. Consider the viral load containment strategy. It can be seen that $\PP(Y_{t_{j+1}}> \eta|A_{t_j}=0, Y_{t_j},\bar A_{t_{j-1}},C)=1-\Phi(\frac{\eta-Y_{t_{j+1}}-\mu_{1}^*(t_{j+1}-t_j)-\gamma^*_C(t_{j+1}-t_j)C}{\tau^* \sqrt{(t_{j+1}-t_j)}})$, where $\Phi(.)$ is the cdf of the standard normal distribution. Thus the $A_{t_j}$, and hence the risk function, are non-linear functions of the random effect and of $C$, leading to different conditional and marginal effects.
Note also that the adaptive strategies give in general different treatment regimes for different subjects, leading to ``personalized medicine''; this will be developed in Section \ref{estphyslaw}.

\section{Designing strategies}\label{sec:strategies}
\subsection{Optimal strategies}\label{sec:optstrategies}
Knowledge of the physical law of $Y$ can be used to design adaptive interventions, that is, strategies.
Once the risk function is chosen one may try to find the best strategy, that is one which minimizes this risk.
This is a special case of optimal control. Optimal control has been well developed, in particular in automatics \citep{bertsekas1995dynamic}; an application for  optimizing antiretroviral treatment has been given by \citet{kirschner1997optimal}.

However, there are several difficulties for a realistic application of optimal control. The first difficulty is that the treatment strategies are restricted in practice. One of the most important restriction is that the treatment can be changed only at discrete visit-times. The second difficulty is that optimal control  may be computationally intractable if there are an even moderately large number of visit-times. A third difficulty is that the ``physical law'' is unknown in practice, so it must be updated at each visit time: see Section \ref{condlawunknown}. Thus, optimal control is most often impractical in this context.

\subsection{Optimal parametric strategies}\label{sec:optparamstrategies}

A possibility is to restrict the space of strategies to parametric forms.
For instance a parametric family of strategies is:
\begin{equation} {\rm logit} [\PP(A_{it_{ij}}=1|\bar Z_{ij},\bar A_{it_{ij}-1}, C_i)] =\alpha_0+\alpha_Z Z_{ij}+\alpha_C C_i\label{eq:paramstrat}.\end{equation}
Note that strategies do not need to be random and deterministic strategies make more sense for clinical use. A parametric family of deterministic strategies is:
\begin{equation}A_{it_{ij}}=1_{\{ Z_{ij}>\beta_0+\beta_C C_i\}}.\label{eq:paramstratdet}\end{equation}


If we are able to compute $R_m^{\intr} $ for all strategies belonging to a parametric class, we can use a minimization algorithm to choose the best strategy in that class.

In fact we do not need to restrict to strategies linear in $C$. Especially if $C$ is continuous we can personalize the strategy by defining it as:
\begin{equation}A_{it_{ij}}=1_{\{ Z_{ij}>\beta_{i}\}}.\label{eq:paramstratdetpers}\end{equation}
If $C$ is binary (\ref{eq:paramstratdet}) and (\ref{eq:paramstratdetpers}) are equivalent, otherwise  (\ref{eq:paramstratdetpers}) is less restrictive.

\subsection{Prediction-based strategies}\label{sec:suboptstrategies}

A practical approach to adaptive regimes is to possibly change the treatment at visit-times in a way to optimize a risk function defined at a not too far future and not depending on future treatment, and updating the risk function at each visit-time (rather than defining it once for all). In our example, we can compute the distribution of $Y_{it_{ij}+1}$ given $(A_{it_j}, \bar Z_{ij},\bar A_{t_{ij}-1},C_i)$ which allows computing any risk function depending on this distribution and of the treatment given at $t_{ij}$. In our example where $Y$ is the viral load, a possible risk function is based on $\PP(Y_{it_{ij}+1}> \eta|A_{it_j}, \bar Z_{ij},\bar A_{it_{ij}-1},C_i)$, if it is admitted that under a certain level, the infection is controlled; for instance one speaks of virological failure for HIV infected patients if the viral load is above the detection limit of 50 copies per ml. The rule could then be:
\begin{equation} A_{it_{ij}}=1_{\PP(Y_{it_{ij}+1}> \eta|A_{it_{ij}}=0, \bar Z_{ij},\bar A_{t_{ij}-1},C_i)> \kappa}, \label{PredbasedStrat} \end{equation}
where $\kappa$ is a small number, say $0.05$: in words, if it is not very likely that without treatment the viral load will remain under the threshold $\eta$, then give the treatment. We shall call this strategy ``the viral load containment strategy".

A more realistic example was given in \citet{prague2012treatment} where the dose of antiviral treatment can be changed at each visit-time to optimize a risk function defined at the next visit-time; see Section \ref{condlawunknown}.

Another type of strategies is to optimize the next visit-time. An example is given in the work of \citet{Villain2018} where the next visit time for IL7 injection is decided as a function of previous observations. More generally, we can define both the updating of the treatment and the next visit time.

\subsubsection{Combining prediction-based and parametric strategies}\label{sec:combiningstrategies}
It is possible to include predictions into parametric families of strategies, leading possibly to better strategies.
For instance we could consider the deterministic parametric strategy obtained from Equation (\ref{PredbasedStrat}) by parameterizing $\kappa$:

\begin{equation}A_{it_{ij}}=1_{\{ \PP(Y_{it_{ij}+1}> \eta|A_{it_j}=0, \bar Z_{ij},\bar A_{t_{ij}-1},C_i)>\beta_{i}\}}.\label{eq:combstratdet}\end{equation}

\subsection{Computation of risk functions for a parametric strategy}\label{sec:comprisk}

For a non-random treatment regime the risk is easy to compute. For strategies, however, it is difficult, or impossible, to analytically solve the stochastic differential equations defined by the system, the observation and the strategy.
The risk can still be computed by simulation.
The conditional risk must be computed for a particular patient for whom we observe the value of $C$, say $c$; here again we drop the subscript $i$. The strategy is specified by a parameter $\beta$ such as in Equations (\ref{eq:paramstratdetpers}) or (\ref{eq:combstratdet}). So, at time $t_0$ we observe $C=c$ and $Z_0$, and we wish to compute the conditional risk $R^{\intr}_c$.
For system and observation equations (4) and (5) the algorithm can be as follows:
For $j=0$ to $j=J$:
\begin{enumerate}
    \item at time $t_j$ we apply the strategy to choose the value of $A_{t_j}$;
\item Conditional on $Y_{t_{j}}$ $Y_{t_{j+1}}$ has the distribution $\NN [Y_{t_{j}}+(\mu_1^* + \gamma^*_C c+ \gamma^*_A A_{t_j})(t_{j+1}-t_{j}), \tau^{*2}(t_{j+1}-t_{j})]$; draw a value from this distribution;
\item generate a value for $Z_{j+1}$ by drawing from $\NN (Y_{t_{j+1}}, \sigma^{*2}_{\varepsilon})$ and go to step 1
\end{enumerate}
This is done for instance $K=1000$ times
 and then one can compute an approximate version of risk (\ref{additiverisk-seuil}) by $\frac{1}{JK}\sum_{k=1}^K\sum_{j=1}^J(1_{ Y^k_{t_j}>\eta}+ \omega A^k_{t_j})$.
We call this algorithm of simulation for known parameters the SKP algorithm.

\section{Inference}\label{condlawunknown}
\subsection{Estimation of the law}
If we have a well-specified model for the  law of $Y$, and if we have enough observations from an observational study, we can estimate this law, for instance by maximum likelihood. In particular we can estimate the compensator of $Y$, $\lambda^{\FF}_Y$, in the filtration $\FF$. This can be done without modeling the law of $A$ and $C$ if these processes are observed \citep{commenges2015stochastic}. Then the estimates of the densities can be plugged in the formulas for the risk to obtain estimates of the risks of different interventions. A Bayesian approach would allow taking into account the uncertainty about the parameters by integrating them out.

As an example, dynamical models have been proposed by \citet{prague2016dynamic} and marginal effects computed by simulation.

\subsection{Observe the marginal law directly by experimentation}
The influence graph of a simple experiment is displayed in Figure \ref{graph-causal-obs} (b); it is similar to the ``do operator'' of \citet{Pearl2000} and this approach has been formalized by \citet{arjas2012causal}. In this experimental setting, the law of the three variables $Y,A,C$ is given by a probability measure $\PP^{ex:A=a}$; this is a particular type of intervention where $C$ does not influence $A$, that is $f^{ex}_{A|C}=f^{ex}_{A}$. If we have observations from an experiment for different values of $a$,  we can estimate the marginal distribution of $Z$ given $A$, and hence that of $Y$, given $A$. The simplest experiment uses fixed values of $A$: $A_{t_j}=a$, and $N/2$ subjects have value $a=1$, $N/2$ subjects have value $a=0$; the values are attributed by randomization, which ensures that they are not influenced by any confounding factor. However, nothing prevents doing an experiment for comparing two (or more) adaptive strategies; this allows directly estimating the marginal risk for the interventions tried.

The great advantage of experiments is that we observe directly the marginal law of $Z$ given $A$. Thus we are free from the NUC assumption. However, there are several important limitations (feasibility, selected subjects, short follow-up, non-compliance), which implies that we still need to analyze observational data. The other advantage is that it is not necessary to estimate the physical law if we are only interested in the marginal risk.

\subsection{Estimate the physical law then make estimation for any intervention}\label{estphyslaw}
In our example, assuming that the true values for subject $i$ are given by Equation (\ref{eq:diffDoobidv}), a well specified model for the observations may be:
\begin{equation}  Z_{ij}=\mu_{0i}+(\mu_{1i} + \gamma_{C} C) t_{ij}+ \gamma_A \int_0^{t_{ij}} A_{iu} \dd u + \tau B_{it_{ij}}+\varepsilon_{ij} ~~j=1,J~~;~~ i=1,N,  \label{eq:obsidv} \end{equation}
with $\mu_{0}\in \Re$,  $\mu_{1}\in \Re$,  $\gamma_{C}\in \Re$, $\gamma_{A}\in \Re$, $\tau \in \Re^+$ $\sigma_{\varepsilon} \in \Re^+$, and with $b_{0i}$ and $b_{1i}$ normal random effects with standard deviations $\sigma_{\mu_0}\in \Re^+$ and  $\sigma_{\mu_1}\in \Re^+$, respectively. Parameters for this mixed linear model can be estimated by maximum likelihood from observational data, as is conventional. Note that the estimation uses a probability conditional on $A$ so that there is no need to model the dynamics of $A$. Once we have estimated the parameters of the physical law of $Y$, we can estimate the conditional and marginal effects of any intervention.

\subsection{Optimizing strategies when the parameters are unknown}
The simplest way is to plug-in the maximum likelihood estimates into the computation of the risk by the SKP algorithm.

A more elaborate approach is to take the uncertainty about parameters into account via a Bayesian approach. This has been done for prediction-based strategies by \citet{prague2012treatment} and \citet{Villain2018}. A way to extend the optimal parametric strategies so as to take parameter uncertainty into account is to include this uncertainty in the computation of the risk.
If there were no random effect the algorithm can be a variation of the SKP algorithm where we include in the simulation a drawing of the parameter values from their posterior distribution; we call it simulation from posterior distribution of parameters (SPDP).

Moreover, there is the possibility of taking into account the cumulative information that we collect on the random parameters. The threshold $\beta$ that specifies the decision rule should then change at each visit time where more information is available. As a consequence we cannot estimate the risk for each fixed value of $\beta$ and then optimize once for all, but we have to optimize at each visit time. This leads not simply to a change of algorithm but to a change of strategy. We call this strategy ``dynamic threshold decision rule (DTDR).

The algorithm can be as follows:
For $j=0$ to $j=J$:
\begin{enumerate}
\item compute the posterior distribution of the random parameters based on observations up to $t_j$;
    \item at time $t_j$ optimize $\beta_j$ using the SPDP algorithm; let $\beta_j^*$ the optimum value;
    \item apply decision rule with the threshold $\beta_j^*$ to define $A_{t_j}$;
    \item observe  $Z_{j+1}$ and go to step 1;
\end{enumerate}

The posterior distribution can be computed by an MCMC algorithm as in \citet{prague2012treatment} or by a Laplace approximation \citep{rue2009approximate} or simply by using the asymptotic distribution of the penalized maximum likelihood estimators which is justified by the Bernstein-von Mises Theorem \citep{VanderVaart2004} and was used by \citet{Prague2013NIMROD}.

\section{Illustration}\label{sec:illustration}
\subsection{Simulated data}
\begin{figure}[p!]
\centering
\includegraphics[width=0.9\textwidth]{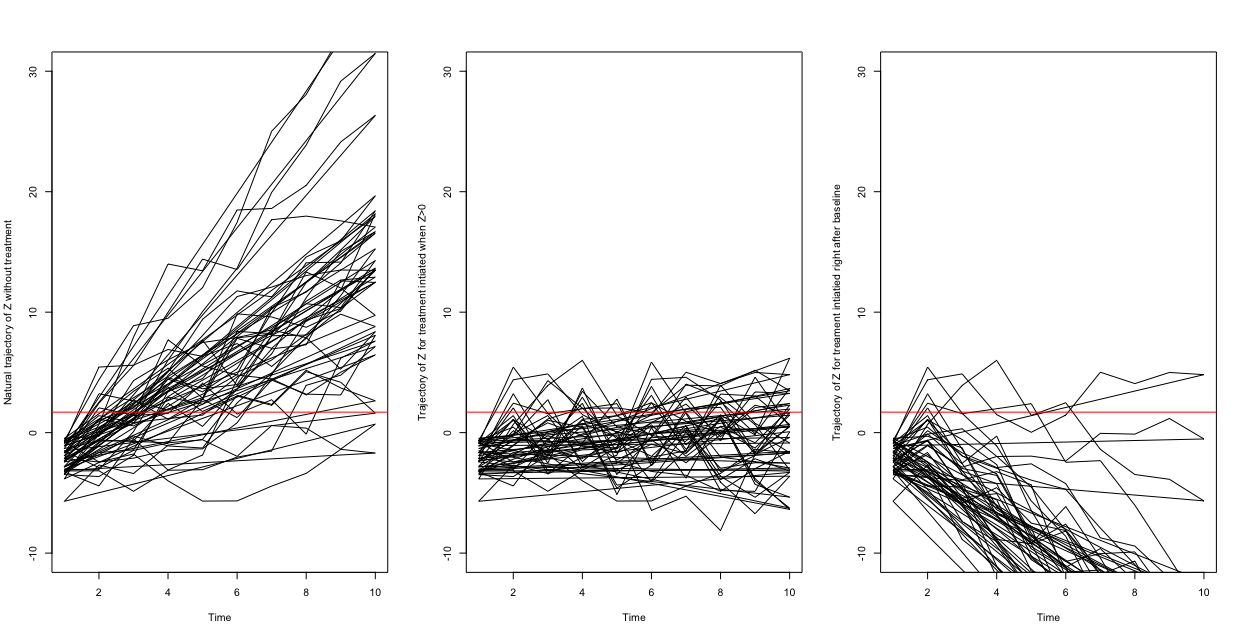}
\caption{Simulated trajectories of biomarker $Z$ for N=30 individuals. (Left) Natural history without initiation of treatment; (Middle) Trajectory of Z when treatment is initiated when  $Z_j>0$; (Right) Trajectory of Z when treatment is initiated right after baseline for all patients. Horizontal red line materializes the  threshold $\eta=1.7$ used in the risk function. \label{fig:datasimul}}
\end{figure}

We simulated observations of samples of N subjects from a system obeying a dynamics similar to Equation (\ref{eq:diffDoobidv}) except that we had both a continuous variable $C$ and a discrete variable $D$:
\begin{equation} \dd Y_{it}=(\mu_{1i}^* + \gamma^*_C C_i+ \gamma^*_D D_i+ \gamma^*_A A_{it})\dd t + \tau^*\dd B_{it}. \label{eq:illu} \end{equation}
  with $Y_{0i} \sim \mathcal{N}(\mu^*_0,\sigma^{*2}_{\mu_0})$, $\mu^*_{1i} \sim \mathcal{N}(\mu_1^*,\sigma^{*2}_{\mu_1})$ and $\epsilon_{ij}\sim \mathcal{N}(0,\sigma^{*2}_\varepsilon)$. $C_i \sim \mathcal{N}(0,1)$ and $D_i\sim \mathcal{B}(0.6)$ were, respectively, continuous and  binary observed covariates.
 \noindent
 We set $\mu^*_0=-2$, $\mu^*_{1}=1$, $\gamma_C^*=0.3$, $\gamma_D^*=1$, $\gamma^*_A=-3$ and $\tau^*=2$.
 The parameters $\sigma^*_{\mu_0}$ and $\sigma^*_{\mu_1}$ depend on the setting of simulation and on whether we assume the individual parameters known or estimated.
 \noindent The observation equation was \begin{equation} Z_{ij}=Y_{t_{ij}}+\varepsilon_{ij}\label{eq:illu-obs}, \end{equation} $i=1,N, j=1,J$ with $J=10$,  and $\sigma_{\varepsilon}^* =0.5$. A well-specified model for the observations $Z_{ij}$ is similar to  Equation (\ref{eq:obsidv}), except for the addition of the variable $D$. Figure \ref{fig:datasimul}  shows the data generated for $30$ subjects under three scenarios: treatment  never initiated,  treatment initiated when the observed Z is higher than 0, treatment initiated right after baseline for all patients.

 We simulated data as they could be found in an observational study of HIV infected patient. In real life, treatment is initiated depending on multiple factors including patients covariates and observed biomarker values, and once initiated, the treatment is most of the time not interrupted. For the simulation, we used a logistic regression such that for patient $i=1,\dots, N$ and visit numbers $j=1,\dots, J$, ${\rm logit}[P(A_{ij}=1|C_i,D_i,Z_{ij})]=-3+2Z_{ij}+0.3C_i+0.5D_i$. The visit-times were taken as $t_{ij}=j$ so that the the last visit time was $t_{iJ}=J$. This model leads to 66.7\% of patients-time under treatment.

\subsection{Estimation of parameters from simulated data}

The parameters, including the causal parameter $\gamma_A$, can be estimated from a real data set by maximum likelihood (ML).
We estimated the parameters of our simulated model using the R package \textit{covBM} which allows to find ML estimates for a mixed linear model with Brownian motion \citep{Stirrup2017} using the function \textit{lmeBM}. Code is available in supplementary material. We performed 1000 replicated estimations on datasets including N=1000 patients. Table \ref{tab:resestimation} summarizes the results and shows that there is essentially no bias in estimation of the parameters of the model, empirical and estimated standard deviation are overall similar and coverage are kept near their nominal value of 95\% (in accordance with the ML theory for well specified models). Overall, we confirm that the causal parameter can  be estimated by maximum likelihood without needing a treatment attribution model assuming model for (\ref{eq:illu}) and (\ref{eq:illu-obs}) is well specified.

\begin{table}[p!]
\hspace{-2cm}
\begin{center}
\begin{tabular}{lcccc}
\hline
Parameters & Bias  & Sd. & Bootstrap Sd. & coverage \\
\hline
$\mu_{0}$& 0.0015 & 0.0352 & 0.0357 & 94.1\%\\
$\mu_{1}$& 0.0012 & 0.0490 & 0.0500 & 94.8\%\\
$\gamma_C$& 0.0006& 0.0266 & 0.0272 & 94.8\%\\
$\gamma_D$& 0.0003 & 0.0547 &0.0559 & 94.9\%\\
$\gamma_A$& 0.0011 & 0.0486 & 0.0539 & 94.5\%\\
\hline
$\sigma_{\mu_0}$& 0.0004 & - & 0.0339 & 94.3\%\\
$\sigma_{\mu_1}$& 0.0003 & - & 0.0373 & 94.9\%\\
\hline
$\tau$ & 0.0008 & - & 0.0286 & 95.0\%\\
$\sigma_\epsilon$& 0.0045 & - &0.0466 & 95.1\%\\
\hline
\end{tabular}
\end{center}
\caption{Simulation on 1000 replicates for evaluating the performances of the estimation from observational data using the function \textit{lmeBM} of package \textit{covBM}.}
\label{tab:resestimation}
\end{table}

\subsection{Simulation of optimal strategy for a given individual assuming that his parameters are known}

In this section we examine strategies where treatment can be temporarily interrupted, and we evaluate them by an approximate version of the risk function (\ref{additiverisk-seuil}) with $\eta = 1.7 \approx \log_{10} (50)$. We will examine parametric strategies given by Equation (\ref{eq:paramstratdetpers}). The parameters of the model are supposed known: here, we set the values to the one used for the simulations. We investigate the optimal strategy for 4 different patients: $(D_1=1, C_1=0.5)$, $(D_2=1, C_2=-0.5)$, $(D_3=0, C_3=0.5)$, $(D_4=0, C_4=-0.5)$. We assumed the parameters  for these patients perfectly known so that $\sigma^*_{\mu_0}=\sigma^*_{\mu_1}=0$.  Table \ref{tab:SKP} summarizes the results for multiple values of $\omega$. We observe that the larger $\omega$, the higher the threshold $\beta$ for treatment initiation. The overall cost $R_{m}^{int}$ is an increasing function of $\omega$. Finally, we observe as expected that for any given $\omega$ the overall cost $R_{m}^{int}$ is higher for individual with steeper increase in their biomarker $Y$, see comparison for patient $(D_1=1, C_1=0.5)$ versus $(D_4=0, C_4=-0.5)$. The strategy is such that for lower $\omega$ the optimal strategy is to treat a lot of patient. When the treatment burden gets higher, i.e. $\omega$ is larger, the strategy progressively tends to ``never treat''.

\begin{landscape}
\begin{table}[p!]
\hspace{-2cm}
\begin{center}
\begin{tabular}{lcccccccccc}
\hline
& \multicolumn{5}{c}{$D_i=0$}& \multicolumn{5}{c}{$D_i=1$} \\\cline{2-11}
$\omega$  & $\beta_1$ & Cost Y & Cost Trt & $R_m^{int}$ &\% Treated & $\beta_1$ & Cost Y & Cost Trt & $R_m^{int}$ &  \% Treated \\
\hline
\multicolumn{5}{l}{$C_i=0.5$}\\
  0.000 & -9.058 & 0.015 & 0.000 & 0.015 & 0.558 &   -11.565 & 0.052 & 0.000 & 0.058 & 0.758 \\
   0.100 & -4.040 & 0.016 & 0.038 & 0.054 & 0.422 &   -5.681 & 0.053 & 0.062 & 0.128 & 0.624 \\
   0.200 & -3.131 & 0.019 & 0.072 & 0.090 & 0.397 &   -4.304 & 0.058 & 0.118 & 0.195 & 0.590 \\
   0.300 & -2.715 & 0.021 & 0.104 & 0.126 & 0.387 &   -3.934 & 0.060 & 0.174 & 0.261 & 0.581 \\
   0.400 & -2.708 & 0.022 & 0.139 & 0.161 & 0.387 &   -3.637 & 0.063 & 0.230 & 0.325 & 0.574 \\
   0.500 & -2.362 & 0.024 & 0.169 & 0.193 & 0.376 &   -3.584 & 0.064 & 0.286 & 0.389 & 0.572 \\
   0.600 & -2.056 & 0.028 & 0.198 & 0.226 & 0.367 &   -3.590 & 0.063 & 0.344 & 0.452 & 0.573 \\
   0.700 & -1.875 & 0.032 & 0.227 & 0.259 & 0.361 &   -3.384 & 0.066 & 0.397 & 0.515 & 0.568 \\
   0.800 & -1.560 & 0.039 & 0.254 & 0.292 & 0.352 &   -3.037 & 0.072 & 0.447 & 0.576 & 0.559 \\
   0.900 & -1.405 & 0.043 & 0.281 & 0.324 & 0.347 &   -3.039 & 0.072 & 0.503 & 0.639 & 0.559 \\
   1.000 & -1.273 & 0.047 & 0.308 & 0.355 & 0.342 &   -2.703 & 0.079 & 0.552 & 0.701 & 0.552 \\
   1.500 & -1.148 & 0.051 & 0.458 & 0.509 & 0.339 &   38.308 & 0.737 & 0.000 & 0.819 & 0.000 \\
   3.000 & 31.284 & 0.546 & 0.000 & 0.546 & 0.000 &   38.308 & 0.737 & 0.000 & 0.819 & 0.000 \\
\hline
\multicolumn{5}{l}{$C_i=-0.5$}\\
   0.000 & -7.508 & 0.010 & 0.000 & 0.010 & 0.450 & -11.565 & 0.037 & 0.000 & 0.037 & 0.781  \\
   0.100 & -3.313 & 0.012 & 0.030 & 0.042 & 0.336 & -4.154 & 0.040 & 0.052 & 0.092 & 0.583   \\
   0.200 & -3.015 & 0.013 & 0.059 & 0.072 & 0.328 & -3.561 & 0.043 & 0.102 & 0.145 & 0.566   \\
   0.300 & -2.762 & 0.015 & 0.087 & 0.101 & 0.321 & -3.427 & 0.043 & 0.152 & 0.195 & 0.562   \\
   0.400 & -2.175 & 0.018 & 0.110 & 0.129 & 0.307 & -3.306 & 0.044 & 0.201 & 0.246 & 0.560   \\
   0.500 & -1.972 & 0.020 & 0.135 & 0.155 & 0.301 & -3.388 & 0.044 & 0.253 & 0.296 & 0.561   \\
   0.600 & -1.903 & 0.021 & 0.161 & 0.182 & 0.299 & -3.036 & 0.048 & 0.298 & 0.346 & 0.551   \\
   0.700 & -1.712 & 0.023 & 0.186 & 0.209 & 0.295 & -2.973 & 0.050 & 0.346 & 0.395 & 0.549   \\
   0.800 & -1.434 & 0.028 & 0.207 & 0.235 & 0.288 & -2.812 & 0.052 & 0.393 & 0.445 & 0.546   \\ 0 & 0.260 & 0.272 & -2.548 & 0.057 & 0.436 & 0.493 & 0.538 &  \\
   1.000 & -0.872 & 0.040 & 0.244 & 0.284 & 0.271 & -2.517 & 0.057 & 0.484 & 0.541 & 0.538   \\
   1.500 & -0.264 & 0.064 & 0.343 & 0.407 & 0.254 & 38.433 & 0.697 & 0.000 & 0.697 & 0.000   \\
   3.000 & 31.284 & 0.446 & 0.000 & 0.446 & 0.000 & 38.433 & 0.697 & 0.000 & 0.697 & 0.000   \\
\hline
\end{tabular}
\end{center}
\caption{Results of the optimal parametric SKP strategies for four individuals: $(D_1=1, C_1=0.5)$, $(D_2=1, C_2=-0.5)$, $(D_3=0, C_3=0.5)$, $(D_4=0, C_4=-0.5)$, for different values of $\omega$.}
\label{tab:SKP}
\end{table}
\end{landscape}

\subsection{Simulation of optimal strategy for a given individual assuming that his individual parameters are unknown}

\begin{figure}[p!]
\centering
\includegraphics[width=0.9\textwidth]{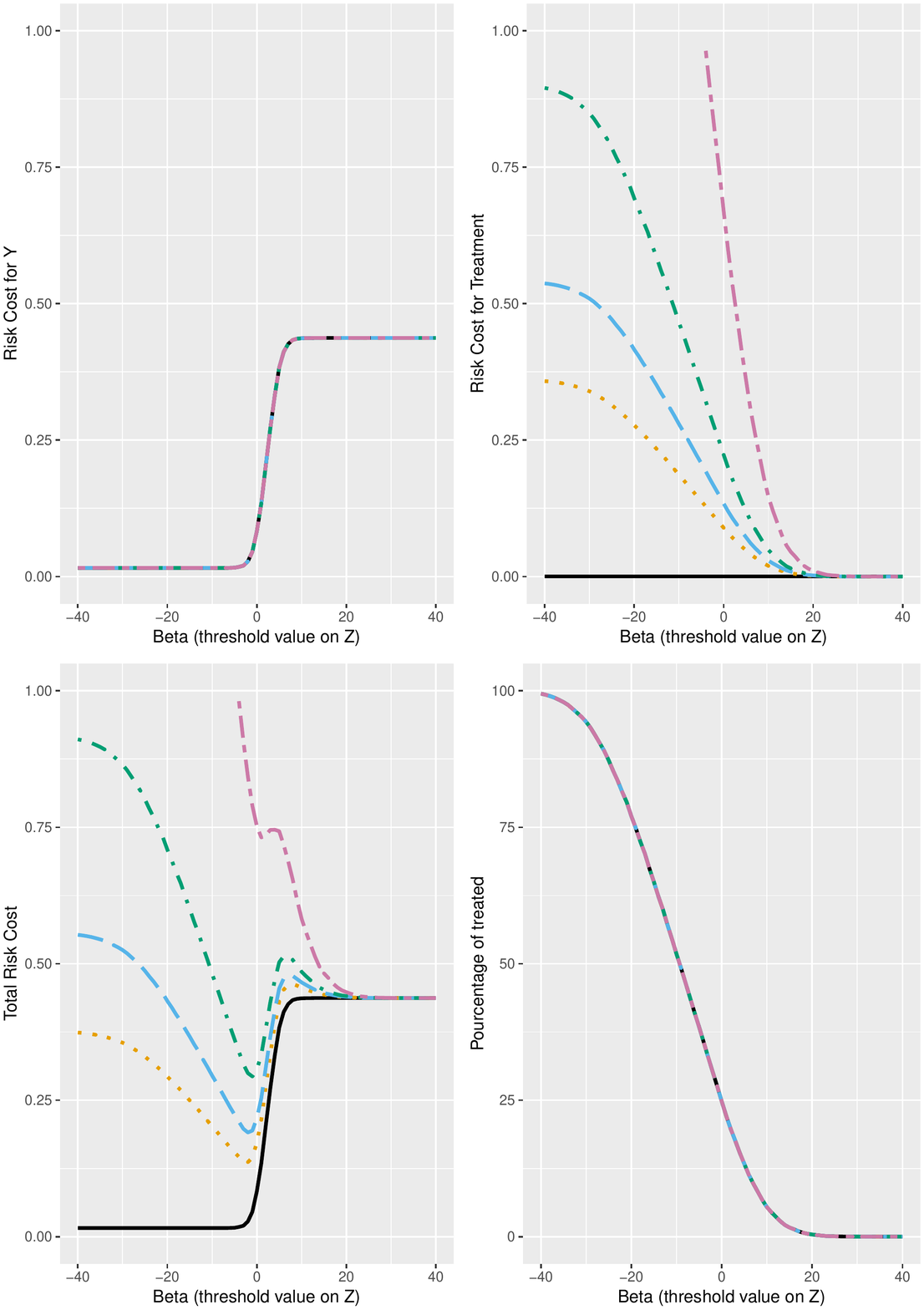}
\caption{ Trajectories of risks for Y, risks for treatment, overall risk and percentage of time under treatment for patient 2 ($D_4=0, C_4=-0.5$) depending on $\beta_1$ the chosen threshold value for $Z(t)$ leading to treatment initiation for multiple values of $\omega$: $\omega=0$ in solid black, $\omega=0.4$ in dotted orange, $\omega=0.6$ in dashed blue, $\omega=1$ in dotted-dashed green and $\omega=3$ in two-dashed pink line. \label{fig:optimrisk}}
\end{figure}

\begin{figure}[p!]
\centering
\includegraphics[width=0.9\textwidth]{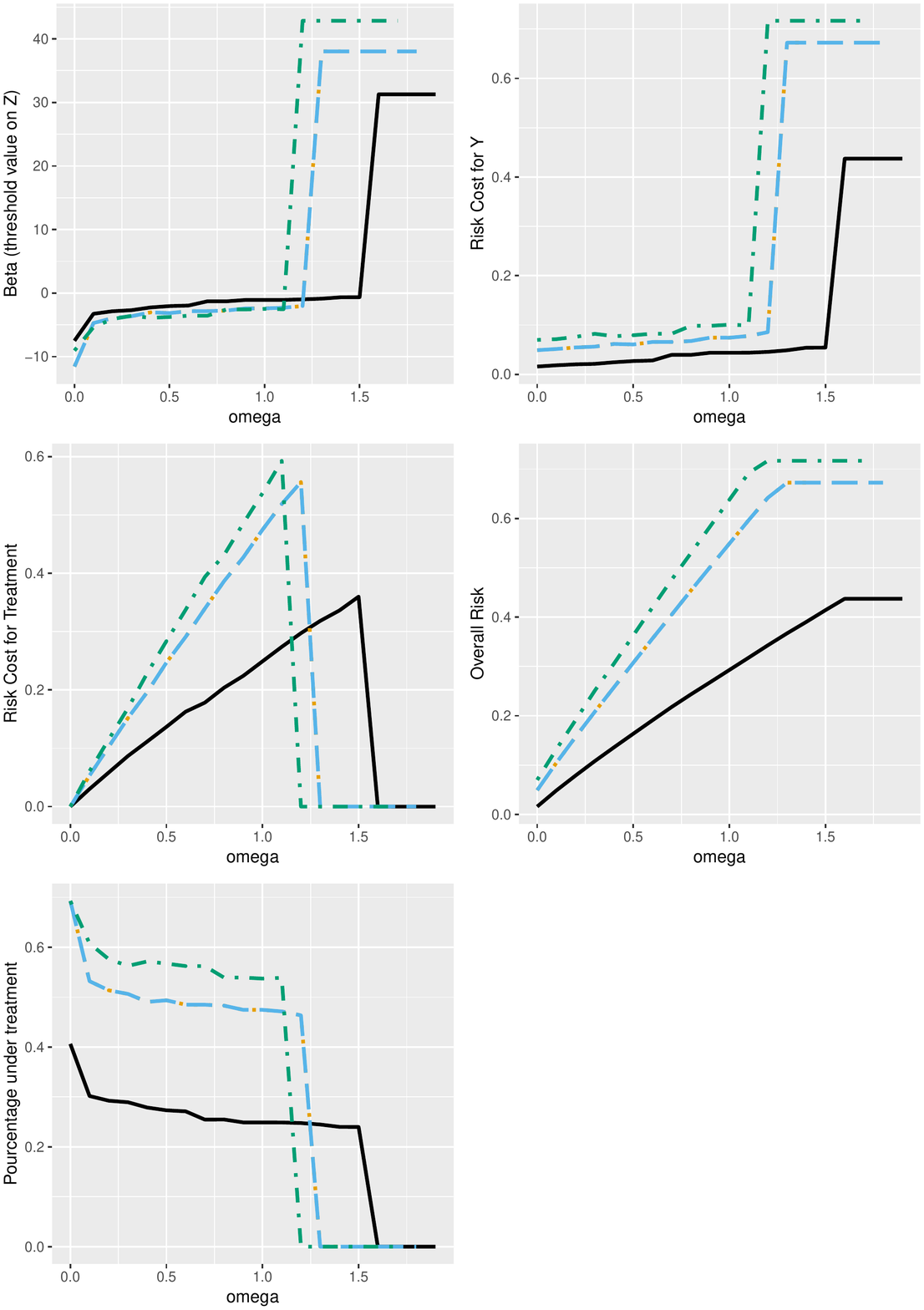}
\caption{Optimal value of threshold $\beta_1$, optimal risk associated to Z, associated to treatment and overall, and percentage of treatment attribution depending on value of $\omega$ for 4 patients ($(D_1=1, C_1=0.5)$ , $(D_2=1, C_2=-0.5)$, $(D_3=0, C_3=0.5)$, $(D_4=0, C_4=-0.5)$).  \label{fig:omegaseuil}}
\end{figure}

In this section we assume that we know the population parameters of the model, but that the parameters with random effects are only known with uncertainty; that is we know that $\mu_{0i}$ and $\mu_{1i}$ come from a normal distribution with known expectation and standard deviations $\sigma^*_{\mu_0}=1$ and $\sigma^*_{\mu_1}=0.5$, respectively. Thus this is an illustration of the SPDP algorithm. Table \ref{tab:SPDP}  summarizes the results for multiple values of $\omega$. It is very similar to results obtained with the SKP algorithm presented in table \ref{tab:SKP}. Figure \ref{fig:optimrisk} shows for patient 4 $(D_4=0, C_4=-0.5)$ the risk function for multiple values of $\beta_1$ according to the value of $\omega$. We see that the overall cost $R_m^{int}$ to optimize has a local and a global minimum for average values of $\omega$. As expected, the cost in $Y$ is an increasing function of the threshold $\beta_1$ and the cost in treatment and the percentage of individual under treatment is a decreasing function of the threshold $\beta_1$. Figure \ref{fig:omegaseuil} shows for the four patients ($(D_1=1, C_1=0.5)$, $(D_2=1, C_2=-0.5)$, $(D_3=0, C_3=0.5)$, $(D_4=0, C_4=-0.5)$) the threshold value $\beta_1$ for treatment initiation, the multiple risks and the percentage of individual under treatment, according to various values of omega. We notice that the overall risk $R_m^{int}$ is a continuous function of $\omega$ suggesting that there is a threshold in $\omega$ from which it is better not to treat at all rather. We find again that optimal $R_m^{int}$ is higher for individuals with steeper increase in their biomarker $Y$, see comparison for patient $(D_1=1, C_1=0.5)$ versus $(D_4=0, C_4=-0.5)$.


\begin{landscape}
\begin{table}[p!]
\hspace{-2cm}
\begin{center}
\begin{tabular}{lcccccccccc}
\hline
& \multicolumn{5}{c}{$D_i=0$}& \multicolumn{5}{c}{$D_i=1$} \\\cline{2-11}
$\omega$  & $\beta_1$ & Cost Y & Cost Trt & $R_m^{int}$ &\% Treated & $\beta_1$ & Cost Y & Cost Trt & $R_m^{int}$ &  \% Treated \\
\hline
\multicolumn{5}{l}{$C_i=0.5$}\\
  0.000 & -11.565 & 0.023 & 0.000 & 0.023 & 0.621 & -9.058 & 0.070 & 0.000 & 0.070 & 0.770 \\
   0.100 & -3.900 & 0.026 & 0.037 & 0.063 & 0.416 & -5.374 & 0.071 & 0.061 & 0.132 & 0.675 \\
   0.200 & -3.573 & 0.027 & 0.073 & 0.100 & 0.406 & -4.085 & 0.076 & 0.115 & 0.191 & 0.640 \\
   0.300 & -2.918 & 0.031 & 0.105 & 0.136 & 0.389 & -3.582 & 0.082 & 0.169 & 0.251 & 0.625 \\
   0.400 & -2.872 & 0.031 & 0.139 & 0.170 & 0.387 & -3.901 & 0.077 & 0.229 & 0.306 & 0.635 \\
   0.500 & -2.371 & 0.037 & 0.168 & 0.205 & 0.373 & -3.765 & 0.079 & 0.284 & 0.363 & 0.630 \\
   0.600 & -1.992 & 0.041 & 0.196 & 0.237 & 0.363 & -3.566 & 0.082 & 0.337 & 0.420 & 0.625 \\
  0.700 & -1.997 & 0.041 & 0.229 & 0.270 & 0.363 & -3.563 & 0.082 & 0.394 & 0.476 & 0.625 \\
   0.800 & -1.917 & 0.043 & 0.260 & 0.303 & 0.361 & -2.568 & 0.098 & 0.431 & 0.530 & 0.599 \\
   0.900 & -1.559 & 0.050 & 0.284 & 0.334 & 0.351 & -2.578 & 0.098 & 0.485 & 0.584 & 0.599 \\
  1.000 & -1.478 & 0.052 & 0.314 & 0.366 & 0.349 & -2.486 & 0.101 & 0.537 & 0.638 & 0.597 \\
   1.500 & 35.042 & 0.516 & 0.000 & 0.516 & 0.000 & 42.851 & 0.717 & 0.000 & 0.717 & 0.000 \\
      3.000 & 35.042 & 0.516 & 0.000 & 0.516 & 0.000 & 42.851 & 0.717 & 0.000 & 0.717 & 0.000 \\
\hline
\multicolumn{5}{l}{$C_i=-0.5$}\\
   0.000 & -7.508 & 0.016 & 0.000 & 0.016 & 0.452 & -11.565 & 0.049 & 0.000 & 0.049 & 0.770 \\
  0.100 & -3.260 & 0.018 & 0.030 & 0.049 & 0.335 & -4.721 & 0.051 & 0.053 & 0.105 & 0.591 \\
   0.200 & -2.861 & 0.020 & 0.059 & 0.079 & 0.325 & -3.950 & 0.054 & 0.103 & 0.157 & 0.571 \\
  0.300 & -2.693 & 0.021 & 0.087 & 0.108 & 0.322 & -3.646 & 0.056 & 0.152 & 0.208 & 0.563 \\
  0.400 & -2.263 & 0.025 & 0.112 & 0.136 & 0.310 & -3.038 & 0.062 & 0.196 & 0.258 & 0.545 \\
  0.500 & -2.052 & 0.027 & 0.137 & 0.164 & 0.304 & -3.152 & 0.061 & 0.247 & 0.307 & 0.549 \\
  0.600 & -1.966 & 0.028 & 0.163 & 0.191 & 0.301 & -2.848 & 0.066 & 0.291 & 0.357 & 0.539 \\
   0.700 & -1.291 & 0.040 & 0.178 & 0.218 & 0.283 & -2.844 & 0.066 & 0.339 & 0.405 & 0.539 \\
  0.800 & -1.296 & 0.040 & 0.204 & 0.243 & 0.283 & -2.757 & 0.068 & 0.386 & 0.454 & 0.537 \\
  0.900 & -1.084 & 0.044 & 0.224 & 0.268 & 0.277 & -2.417 & 0.074 & 0.427 & 0.502 & 0.527 \\
   1.000 & -1.087 & 0.044 & 0.249 & 0.293 & 0.277 & -2.419 & 0.074 & 0.475 & 0.549 & 0.527 \\
   1.500 & -0.652 & 0.054 & 0.360 & 0.414 & 0.267 & 38.035 & 0.672 & 0.000 & 0.672 & 0.000 \\
 3.000 & 31.284 & 0.437 & 0.000 & 0.437 & 0.000 & 38.035 & 0.672 & 0.000 & 0.673 & 0.000 \\
\hline
\end{tabular}
\end{center}
\caption{Results of the optimal parametric SPDP strategies for four individuals ($(D_1=1, C_1=0.5)$, $(D_2=1, C_2=-0.5)$, $(D_3=0, C_3=0.5)$, $(D_4=0, C_4=-0.5)$) for different values of $\omega$ assuming that their individual parameters have uncertainty such that $\sigma^*_{\mu_0}=1$ and $\sigma^*_{\mu_1}=0.5$.}
\label{tab:SPDP}
\end{table}
\end{landscape}

\subsection{Adaptive optimal}
Now that we validated the properties of both the estimation and the optimal control strategy, it is possible to combine both so that we propose a strategy as described in the DTDR strategy.
In our toy-example example it is easy to compute the likelihood of $\bar Z_j$, which can be used for computing the posterior distribution of the parameters or for approximating it. We note that $\bar Z_j$ being obtained by summing normal variables has itself a normal distribution. So, we have only to compute its expectation and variance matrix. The expectations of the $Z_j$ (the components of $\bar Z_j$) are easily computed from Equation (\ref{eq:obsidv}). The computation of the variance (given the parameters) is also easy because we identify independent variables which are $B_{t_j}$ and the $\varepsilon_j$, and so, the variances are obtained by summing the variances of these variables, that is: $\tau^2 t_j+\sigma_{\varepsilon}^2$; the covariance between $Z_j$ and $Z_{j'}$ comes from the covariance  of the Brownian  which is $\tau^2 \min {t_j,t_{j'}}$.

It is possible to neglect the information brought by the incoming observation of a particular subject on the fixed parameters $\theta$, so that we can update only the distribution of the random parameters, $\mu=(\mu_0,\mu_1)$. In that case, we can write $f_{\mu,\theta|\bar Z_j}=f_{\mu|\bar Z_j,\theta}f_{\theta}$. The priors for $\mu$ and $\theta$, denoted $\NN(\nu_0, \Omega_0)$ and $\NN(\theta_0,\Xi_0)$ respectively, come from the observation of a sample, The likelihood of observation of a subject up to visit $j$, $f_{\bar Z_j|\mu,\theta}$ involves $\theta$ but we neglect the variability due to the uncertainty of $\theta$ in this likelihood and make the computation at $\theta_0$. The expectation of $\bar Z_j$ can be written $A_j\mu+c_j$ so that we have $\bar Z_j\sim \NN(A_j \mu+c_j, \Sigma_j)$; thanks to the linearity  in $\mu$ of the expectation, the posterior of $\mu$ is normal (conjugate prior), so that we have just to examine the argument of the exponential part of the density to determine its expectation and variance which specify the posterior.

Applying the formulas of Appendix \ref{Ap:conjugaterpriors} to $(\bar Z_j-c_j)$ (with $c_0=0$) we find the expectation $\nu_j$ and variance  $\Omega_j$ of the posterior of $\mu$:
$$ \nu_j=\Omega_j[A_j^\top \Sigma_j^{-1}(\bar Z_j-c_j)+ \Omega_0^{-1}\nu_0]~~;~~\Omega_j^{-1}=A_j^\top \Sigma_j^{-1}A_j+\Omega_0^{-1}. $$

\section{Conclusion}\label{sec:conclusion}
We have shown how effects of simple and adaptive interventions could be defined and computed within the stochastic system approach to causality. More over we have shown that it is possible to design optimal parametric strategies. This was illustrated in a simulation study showing that the physical law could be learned from an observational study and that it would be possible to control an intermittent antiretroviral treatment in an adaptive way, based on a dynamical model. The simulation for the adaptive optimal strategy remains to do.
\bibliographystyle{chicago}
\bibliography{causality}

\newpage
\section{Appendix}

\subsection{Computing effects of strategies: an analytical formula}

We  derive an analytical formula in the case where the visit-times are fixed and the loss function depends only on values of $Y$ at visit-times. For simplicity we remove the factor $C$.
To compute any risk function we must compute the marginal distribution of $Y_t$ under the intervention. This can be done by recurrence: knowing the marginal distribution of $(\bar Y_{t_j}, \bar A_{t_{j-1}},\bar Z_{t_j})$, compute the marginal distribution of $(\bar Y_{t_{j+1}}, \bar A_{t_{j}},\bar Z_{t_{j+1}})$. For making this computation, we use:
\begin{itemize}
\item the physical law of $Y$, $f_{Y(t{j+1})|\bar Y(t_j), \bar A_{t_j}}$,
\item the strategy $f^{\intr}_{A_{t_j}|\bar Z_{t_j},\bar A_{t_{j-1}}}$.
 \item the observation equation $f_{Z_{t_{j+1}}|Y_{t_{j+1}}}$.
 \end{itemize}

 {\bf Theorem} The recurrence equation is:
 \begin{equation}\label{recurence}f^{\intr}_{\bar Y_{t_{j+1}}, \bar A_{t_{j}},\bar Z_{t_{j+1}}}=\left [f_{Z_{t{j+1}}|Y_{t_{j+1}}}f_{\bar Y_{t_{j+1}}|\bar Y_{t_{j}}, \bar A_{t_{j}}}f^{\intr}_{A_{t_j}|\bar Z_{t_j},\bar A_{t_{j-1}}}\right ]f^{\intr}_{\bar Y_{t_{j}}, \bar A_{t_{j-1}},\bar Z_{t_{j}}}.\end{equation}

 That is, one goes from the marginal distribution at $t_j$ to that at $t_{j+1}$ by multiplying by the product of the conditional densities of observation, physical law and strategy.

 {\bf Proof: }
  We have that $f^{\intr}_{\bar Y_{t_{j+1}}, \bar A_{t_{j}},\bar Z_{t_{j+1}}}=f^{\intr}_{\bar Y_{t_{j+1}}, \bar A_{t_{j}},\bar Z_{t_{j}}}f_{Z_{t{j+1}}|Y_{t_{j+1}}}$ (because $Z_{t_{j+1}}$ depends only on $Y_{t_{j+1}}$.
Then $f^{\intr}_{\bar Y_{t_{j+1}}, \bar A_{t_{j}},\bar Z_{t_{j}}}=f_{ Y_{t_{j+1}}|\bar Y_{t_{j}} \bar A_{t_{j}}}f^{\intr}_{\bar Y_{t_{j}}, \bar A_{t_{j}},\bar Z_{t_{j}}}$ (because the physical law does not involve the observation $\bar Z_{t_{j}}$).
Finally, we have that $f^{\intr}_{\bar Y_{t_{j}}, \bar A_{t_{j}},\bar Z_{t_{j}}}=f^{\intr}_{A_{t_j}|\bar Z_{t_j},\bar A_{t_{j-1}}}f^{\intr}_{\bar Y_{t_{j}}, \bar A_{t_{j-1}},\bar Z_{t_{j}}}$ (because the strategy does not depend on the true values given the observed values $\bar Z_{t_j}$).

When all the distributions are Gaussian, the computation can be done analytically.

For a  general loss function at horizon $t_{j+1}$, the associated risk is:
$$R^{\intr}= \int L(\bar y_{t_{j+1}}, \bar a_{t_{j}})f^{\intr}_{\bar Y_{t_{j+1}}, \bar A_{t_{j}},\bar Z_{t_{j+1}}}(\bar y_{t_{j+1}}, \bar a_{t_{j}},\bar z_{t_{j+1}})\dd \bar y_{t_{j+1}} \dd \bar a_{t_{j}} \dd \bar z_{t_{j+1}}.$$
Except in the Gaussian case and with linear loss function, this computation will be intractable if there are many visit times; in that case, one must resort to simulation.

\subsection{Conjugate normal priors}\label{Ap:conjugaterpriors}
Let the prior for the parameter vector $\theta$ of dimension $m$ be $\theta \sim \NN(\nu,\Omega_0)$ and the likelihood for the $j$-dimensional observation be $Z\sim \NN(A \theta, \Sigma)$, where $A$ is a $ j \times m$  matrix. Then the posterior is normal, $\NN(\nu_p,\Omega_p)$ and by identification of the quadratic forms we find that
$$\Omega_p^{-1}=A^\top \Sigma^{-1}A+ \Omega^{-1}$$
and $$\nu_p=\Omega_p(A^\top\Sigma^{-1}Z+\Omega_0^{-1} \nu_0)$$

\label{lastpage}

\end{document}